\documentclass[aps,pra,showpacs,twocolumn,superscriptaddress]{revtex4-1}
\usepackage{amsmath}
\usepackage{amsfonts,dsfont}
\usepackage{amssymb}
\usepackage{graphicx}
\usepackage{natbib}
\usepackage[colorlinks=true,linkcolor=blue,urlcolor=black,citecolor=blue]{hyperref}

\begin{document}

\title{Dynamical invariants in non-Markovian quantum state diffusion equation}

\author{Da-Wei Luo}
\affiliation{Beijing Computational Science Research Center, Beijing 100084, China}
\affiliation{Department of Theoretical Physics and History of Science, The Basque Country University (UPV/EHU), PO Box 644, 48080 Bilbao, Spain}
\affiliation{Ikerbasque, Basque Foundation for Science, 48011 Bilbao, Spain}

\author{P. V. Pyshkin}
\affiliation{Beijing Computational Science Research Center, Beijing 100084, China}
\affiliation{Department of Theoretical Physics and History of Science, The Basque Country University (UPV/EHU), PO Box 644, 48080 Bilbao, Spain}
\affiliation{Ikerbasque, Basque Foundation for Science, 48011 Bilbao, Spain}

\author{Chi-Hang Lam}
\affiliation{Department of Applied Physics, Hong Kong Polytechnic University, Hung Hom, Hong Kong, China}

\author{Ting Yu}
\affiliation{Beijing Computational Science Research Center, Beijing 100084, China}
\affiliation{Center for Controlled Quantum Systems and Department of Physics and Engineering Physics, Stevens Institute of Technology, Hoboken, New Jersey 07030, USA}

\author{Hai-Qing Lin}
\affiliation{Beijing Computational Science Research Center, Beijing 100084, China}

\author{J. Q. You}
\affiliation{Beijing Computational Science Research Center, Beijing 100084, China}

\author{Lian-Ao Wu}
\email{lianaowu@gmail.com}
\affiliation{Department of Theoretical Physics and History of Science, The Basque Country University (UPV/EHU), PO Box 644, 48080 Bilbao, Spain}
\affiliation{Ikerbasque, Basque Foundation for Science, 48011 Bilbao, Spain}

\date{\today}

\begin{abstract}
We find dynamical invariants for open quantum systems described by the non-Markovian quantum state diffusion (QSD) equation. In stark contrast to closed systems where the dynamical invariant can be identical to the system density operator, these dynamical invariants no longer share the equation of motion for the density operator. Moreover, the invariants obtained {with} bi-orthonormal basis can be used to render an exact solution to the QSD equation and the corresponding non-Markovian dynamics without using master equations or numerical simulations. Significantly we show that we can apply these dynamical invariants to reverse-engineering a Hamiltonian that is capable of driving the system to the target state, providing a novel way to design control strategy for open quantum systems.
\end{abstract}


\maketitle

\section{Introduction}
The theory of open quantum system~\cite{Breuer2002} provides a realistic and complete description that takes into account the often uncontrollable and inevitable interaction between the system under consideration and its environment. This particular field has attracted high attention of physicists because environment-induced effects play a vital role in a wide variety of research topics such as quantum information~\cite{Nielsen2000}, quantum transport~\cite{Rebentrost2009} and quantum optics~\cite{Scully1997}. Indeed, in practical quantum information processing, the inevitable interactions between the system and the environment generally lead to a deterioration of quantum information which is one of the biggest hurdle of building quantum devices or setups. Conventionally, the Markov approximation was extensively used because of its simplicity {and validity for systems where the system-bath coupling is weak and the memory effect of the bath is neglectable}. The Markovian approximation entails that the open system dynamics is forgetful and is valid only when memory effects of the environment are negligible. However, this approximation becomes invalid when the system-environment coupling is strong or when the environment is structured~\cite{Breuer2002}. Consequently, general non-Markovian environments have to be considered in explaining new experimental advances in quantum optics~{\cite{exp_a1}}, as well as in various quantum information tasks where environmental memory can be utilized to control entanglement dynamics~{\cite{exp_a2}}. Therefore, it is vital to have a non-Markovian description of the system's dynamics under the influence of the memory effects and the back-action of the environment without making any approximation. However, a precise description of non-Markovian open systems has long been a challenge. To this end, many theoretical approaches have been developed~\cite{Bellomo2007,Piilo2008,Tu2008,Diosi1998,Strunz1999}. Among them, a stochastic Schr\"odinger equation called the non-Markovian quantum state diffusion (QSD)~\cite{Diosi1998,Strunz1999} which was derived from a microscopic Hamiltonian has several advantages over other exact master equations and has been proven to be a powerful tool in study of the system dynamics. While originally derived for systems embedded in bosonic bath, the QSD framework has been extended to deal with fermonic bath as well~\cite{fermionic_Chen2013,Fermionic_Zhao2012}. Exact master equations were derived for many interesting systems such as dissipative multi-level atoms~\cite{Jing2012}, multiple qubits~\cite{Jing2013} and quantum Brownian motion~\cite{Diosi1998,Strunz2004} which was also exactly given via a path-integral approach~\cite{Hu1992}. Recently, a generic tool for deriving non-Markovian master equation has been developed using QSD~\cite{Chen2014} which is applicable to a generic open quantum system irrespective of the system-environment coupling strength and the environment frequency distribution. Quantum continuous measurement~\cite{Diosi2008,Wiseman2008,Wu04} and quantum control method~\cite{Jing2013a} employing the QSD  were also studied.

In quantum mechanics, an invariant of a quantum system remains intact during evolution of the system. The Lewis-Riesenfeld dynamical invariant~\cite{Lewis1967,Lewis1969} which was first introduced to find the solutions of time-dependent Schr\"{o}dinger equations has been used lately to engineer quantum states~\cite{Jing2013b,Chen2011}, perform quantum computation tasks~\cite{Sarandy2011} as well as study shortcuts to adiabaticity~\cite{Ibanez2011}. However, it has been shown that for closed systems under hermitian Hamiltonians, the system density matrix itself (evolved by the propagator) can be a dynamic invariant, since they share the same linear equation of motion. The dynamical invariant has also been extended to non-hermitian Hamiltonians~\cite{Gao1992} and convolutionless master equations~\cite{Sarandy2007}. In this paper, we show that for open systems whose dynamics can be described by the QSD equation, the invariants are no longer equivalent to the reduced density operator. It is also possible to obtain an analytic solution of the QSD equation using the dynamical invariants under a bi-orthonormal basis, yielding new information on the analytical quantum trajectories. Using the QSD invariants, we can also reverse-engineer a Hamiltonian that is capable of driving the system to the target state. Unlike unitary evolutions, this control protocol allows the spectrum of the state to change, making it more appealing to experimental realizations.

\section{QSD equation}
We consider a generic quantum system embedded in a bosonic bath with the Hamiltonian~\cite{Diosi1998,Strunz1999} (setting $\hbar=1$)
\begin{equation}
	H=H_\mathrm{sys}+\sum_k \left(g_k L b_k ^\dagger+g_k^* L^\dagger b_k \right)+\sum_k \omega_k b_k ^\dagger b_k,
\end{equation}
where $H_\mathrm{sys}$ is the Hamiltonian of the system, $L$ is the Lindblad operator, $b_k$ denotes the $k$th-mode annihilation operator of the bosonic bath with frequency $\omega_k$, and $g_k$ stands for the coupling strength. The bath state can be represented by a set of complex numbers $\{z_k\}$ which labels the Bargmann coherent state of each bath mode $k$. One remarkable feature of this open system is that the influence of the bath can be fully encoded in a bath correlation function $\alpha(t,s)=\sum_k |g_k|^2 e^{-i \omega_k (t-s)}$. If we define a function $z_t^*\equiv -i\sum_k g_k^* z_k^* e^{i \omega_k t}$ that characterizes the time-dependent states of the bath and interpret $z_k$ as a Gaussian random variable, then $z_t^*$ becomes a Gaussian random process with a zero mean $\mathcal{M}[z_t^*]=0$ and the correlation function $\alpha(t,s)=\mathcal{M}[ z_t z_s^*]$, where $\mathcal{M}[\ldots]$ stands for the ensemble average. For simplicity, we first consider the case of zero-temperature bath. In this case, the state $|\psi_{z^*}(t) \rangle=\langle z^*|\Psi_{\rm tot}(t)\rangle$, obtained by projecting the total wave function $|\Psi(t)\rangle$ onto the bath state $|z \rangle$, corresponds to a quantum trajectory of the system and obeys a linear, time-local QSD equation~\cite{Diosi1998}
\begin{equation}
	\frac{\partial}{\partial t}|\psi_{z^*}(t) \rangle= \left[-iH_{sys}+Lz^*-L ^\dagger \bar{O}(t,z)\right]|\psi_{z^*}(t) \rangle,\label{qsd}
\end{equation}
where $O$ is an operator ansatz defined by the functional derivative $\frac{\delta}{\delta z_s^*}|\psi_{z^*}(t) \rangle=O(t,s,z^*)|\psi_{z^*}(t) \rangle$, and $\bar{O}(t,z^*)=\int_0^t \alpha(t,s) O(t,s,z^*)ds$. The reduced density operator $\rho_s(t)\equiv\mathrm{Tr}_\mathrm{env}|\Psi_\mathrm{tot} \rangle\langle \Psi_\mathrm{tot}|$ can be obtained as $\rho_s(t)=\mathcal{M}[|\psi_{{z}^*}(t) \rangle \langle{\psi}_{{z}}(t)|]$ by the ensemble average of the quantum trajectories under all possible realizations of the noise function and then the corresponding non-Markovian master equations can in principle be derived. The main challenge in the application of the QSD is to derive the functional derivative $O$-operator. This $O$-operator can be exactly obtained for some simple models (see e.g.~\cite{Diosi1998}) or perturbatively derived for more general systems~\cite{Li2014e}. Note that the QSD equation under a specific noise realization $z_t^*$ can be formally interpreted as a Schr\"{o}dinger equation with the non-hermitian effective Hamiltonian
\begin{equation}
 	H_\mathrm{eff}=H_s+iLz^*-iL^\dagger \bar{O}(t,z^*). \label{heff_qsd_def}
\end{equation} 
Below we use the dynamical invariants to analytically solve the QSD equation, which could give us an explicit expression for the reduced density matrix.

\section{Dynamical invariants in an open quantum system}
The Lewis-Riesenfeld dynamical invariant was first developed~\cite{Lewis1967} to study the magnetic-moment series for a charged particle moving nonrelativistically in an electromagnetic field and later generalized~\cite{Lewis1969} to solve the time-dependent Schr\"odinger equation. The invariant $I(t)$ was defined so that its expectation value under any density operator $\rho(t)$ is \emph{time-independent}, i.e., $\frac{\partial}{\partial t}\mathrm{Tr}\left[\rho(t) I(t)\right]\equiv \frac{\partial}{\partial t}\mathcal{I}\equiv0$.

It has been shown that the Lewis-Riesenfeld dynamical invariant is useful in dealing with time-dependent quantum problems, such as quantum computing in continuous time~\cite{Sarandy2011}. In fact, for closed systems whose dynamics is governed by a hermitian Hamiltonian, the Lewis-Riesenfeld dynamical invariant obeys $\frac{\partial}{\partial t}I(t)=-i[H,I(t)]$, i.e. the von Neumann equation and thus shares the same dynamical behavior as the density operator. As a result, with the knowledge of the dynamical invariant, one can know the dynamics of the system under consideration. It is shown~\cite{Jing2013b} that the dynamical invariant $I(t)$, the propagator $U(t)$ and the density operator $\rho(t)$ are mutually equivalent to each other without considering the Lewis-Riesenfeld phase. Indeed, if we let $I(t)=\rho(t)$, then for any unitary propagator $U(t)$, and density operator $\sigma(t)$, we have  $\mathcal{I}=\mathrm{Tr}\left[\rho(0)\sigma(0)\right]$, which is time independent. The propagator $U(t)$ can also be written as $U(t)=\sum_n |\varphi_n(t)\rangle\langle\varphi_n(0)|$, where $|\varphi_n(t)\rangle$ is the instant eigenvector of the dynamical invariant. Thus, it readily follows that if a closed system is initially prepared in one of the eigenvectors of $I(0)$, then it will necessarily evolve to the instant eigenvector of $I(t)$ with the same index at a later time $t$. This property makes the dynamical invariant a valuable tool for both studying the state engineering~\cite{Jing2013b} and calculating the geometric phases.

In contrast, for an open quantum system whose dynamics is determined by the QSD equation, the problem becomes complicated. The reduced density operator under a particular realization of noise function $z$ can be shown to satisfy
\begin{equation}
	\frac{\partial}{\partial t}P_z(t)=i \left[P_z(t) H_{\mathrm{eff}}^\dagger-H_{\mathrm{eff}}P_z(t)\right]\label{drho},
\end{equation}
where $P_z(t)=|\psi_{z^*}(t) \rangle\langle \psi_{z}(t)|$ and $H_\mathrm{eff}$ is the effective time-dependent non-hermitian Hamiltonian given by the QSD in Eq.~\eqref{qsd}. If one directly defines the dynamical invariant as in a hermitian system by imposing $\frac{\partial}{\partial t}\mathrm{Tr}\left[P_z(t) I(t)\right]\equiv 0$, it can be seen that the invariant satisfies $\frac{\partial}{\partial t}I(t)=i \left[I(t)H_{\mathrm{eff}}-H_{\mathrm{eff}}^\dagger I(t)\right]$,
which differs from Eq.~\eqref{drho} unless $H_\mathrm{eff}$ is hermitian. Thus, the dynamical invariant defined this way does not give the reduced density operator of the system under a given noise channel $z$.

On the other hand, various studies~\cite{Gao1992,Ibanez2011} have used the biorthonormal basis to study the dynamical invariants for a non-hermitian system. In such a framework, a complete biorthonormal set of eigenvectors is introduced so that the left and right eigenvectors of the Hamiltonian are given, respectively, by $H_{\mathrm{eff}}|\psi_\lambda(t)\rangle=\lambda(t)|\psi_\lambda(t)\rangle$ and $\langle\tilde{\psi}_\lambda(t)|H_{\mathrm{eff}}=\lambda(t)\langle\tilde{\psi}_\lambda(t)|$, where the orthonormal condition becomes $\langle\tilde{\psi}_\lambda|\psi_\mu \rangle=\delta_{\mu,\lambda}$ and the completeness is $\sum_\lambda |\psi_\lambda \rangle\langle\tilde{\psi}_\lambda|=\mathds{1}$. Note that it should be carefully checked if such an eigen decomposition indeed exists for the system under consideration, because it may not be always so for any non-hermitian Hamiltonian~\cite{Wong1967}. {For the criteria proposed in~\cite{Wong1967}, we have the system Hamiltonian as the self-adjoint part, and the rest should generally be continuous and bounded for real physical scenarios.}. Then, the time evolution is now governed by $i|\dot{\psi}(t)\rangle=H_\mathrm{eff}|\psi(t)\rangle$ and $i\frac{\partial}{\partial t}|\tilde{\psi}(t)\rangle=H_\mathrm{eff} ^\dagger|\tilde{\psi}(t)\rangle$. As a result, the definition $\frac{\partial}{\partial t}\mathrm{Tr}\left[\tilde{P}_z(t) I(t)\right]\equiv 0$, where $\tilde{P}_z(t)=|\psi_{z^*}(t) \rangle\langle \tilde{\psi}_{z}(t)|$, gives
\begin{equation}
	\frac{\partial}{\partial t} I(t)=-i\left[H_{\mathrm{eff}},I(t)\right],\label{i_heff_def}
\end{equation}
which has the same form as in the hermitian case, albeit with a non-hermitian effective Hamiltonian given by the QSD equation~\eqref{qsd}. Remarkably, this newly defined invariant can be used to give an analytical solution to the QSD equation. Since the dynamical invariants for a non-hermitian Hamiltonian is no longer guaranteed to be hermitian, we should use the biorthonormal basis given by the instantaneous eigenvectors of $I(t)$, i.e., $I(t)|\varphi_\mu(t)\rangle=\mu|\varphi_\mu(t)\rangle$, and $\langle\tilde{\varphi}_\mu(t)|I(t)=\mu\langle\tilde{\varphi}_\mu(t)|$, with $\langle\tilde{\varphi}_\mu(t)|\varphi_\nu(t)\rangle=\delta_{\mu,\nu}$. We formally write the general solution to the QSD equation as $|\psi_{z^*}(t)\rangle=\sum_\mu c_\mu(t)|\varphi_\mu(t)\rangle$ and substitute it into the QSD equation. After some algebra {(See Appendix)}, we find that under the biorthonormal basis of the invariants, the QSD equation becomes an effectively uncoupled set of differential equations of the coefficients $c_\mu(t)$, and its solution is given by
\begin{equation}
	c_\mu(t)=c_\mu(0)\exp \left[-\int_0^t d\tau \left(i \langle\widetilde{\varphi}_\mu|H_\mathrm{eff} |\varphi_\mu \rangle+\langle\widetilde{\varphi}_\mu|\dot{\varphi}_\mu\rangle \right)\right].\label{psi_solu}
\end{equation}
This compact, explicit solution to the QSD equation is our central result and it applies to any realizations of the noise $z_t^*$. With $O$ determined, we can analytically predict the quantum trajectory for each realization of the noise $z_t$, which was previously determined numerically using the QSD method. The reduced density operator of the system can be obtained by analytically taking the ensemble average for all realizations of noises via Novikov's theorem~\cite{Chen2014}.

\begin{figure}[t]
	\begin{centering}
		\includegraphics[scale=.3]{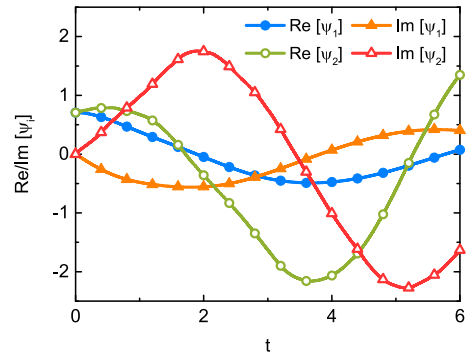}
	\end{centering}
	\caption{(Color online) The real and imaginary part of the wave function versus time $t$ under one random realization of the noise function $z_t$ for the RWA spin-boson model, where we used $\gamma=\Gamma=\lambda=1$. The solid curves correspond to the numerical solutions; the solid, open circles and triangles correspond to the analytical results obtained via the dynamical invariant Eq.~\eqref{i_heff_def}.}
	\label{fig_cf}
\end{figure}

\section{Quantum trajectories}
As an illustrative example, we apply the dynamical invariants method to the quantum dynamics of a dissipative qubit under the rotating-wave approximation (RWA). This model is widely used to display the decoherence effects and is exactly solvable. The Hamiltonian of the system is $H_s=\sigma_z$ and the Lindblad operator is $L=\lambda\sigma_-$, where $\lambda$ is the system-bath coupling strength and $\sigma$'s are Pauli matrices. The $\bar{O}$ operator for this model~\cite{Diosi1998} takes the form of $\bar{O}(t,z^*)=F(t)\sigma_-$, where $F(t)$ is a function depending on the both system parameters and the bath spectral density. We can obtain the dynamical invariant for a given channel $z$ via Eq.~\eqref{i_heff_def} as
\begin{align}
	I(t)&=\sigma_z+2\lambda\int_0^t z^*_u\exp \left[\int_u^t 2i+ \lambda F(s) ds\right] du \sigma_-\nonumber\\
	&\equiv\sigma_z+g(t) \sigma_-.
\end{align}
The left and right eigenvectors can be readily obtained and we finally have
\begin{align}
	|\psi_{z^*}(t)\rangle=&\psi_1(0)\exp \left[-\lambda\int_0^t F(\tau)d\tau-it\right]\begin{bmatrix}1\\\frac{g(t)}{2}\end{bmatrix} \nonumber\\
	&+\psi_2(0)\exp(it)\begin{bmatrix}0\\1\end{bmatrix},
\end{align}
assuming an initial state $|\psi_{z^*}(0)\rangle=[\psi_1(0),\psi_2(0)]^T$, with $T$ denoting the transpose of a matrix. In particular, for the Ornstein-Uhlenbeck noise $\alpha(t,s)=\gamma\Gamma \exp(-\gamma|t-s|)/2$, $F(t)$ can be explicitly given by $\dot{F}(t)=-\gamma F(t)+2iF(t)+\lambda F(t)^2+\lambda\gamma/2$.

Implementation of $\mathcal{M}[|\psi_{z^*}(t)\rangle\langle\psi_z(t)|]$ gives the reduced density operator. This provides us an analytic tool to deal with the QSD equation. For a system with known $O$ operator, we can use it to directly obtain an explicit expression for the reduced density operator as a function of time neither using a master equation nor resorting to numerical calculations. This can be very beneficial for high-dimensional systems whose numerical calculations may be very time-consuming. Another more complex example is the dissipative three-level atom with $H_s=\omega J_z=\omega \left(|0 \rangle\langle 0|-|2 \rangle\langle 2|\right)$ and $L=J_-=\sqrt{2}\left(|0 \rangle\langle 1|+|1 \rangle\langle 2|\right)$. The $\bar{O}$ operator for this model~\cite{Jing_3lvl} explicitly depends on the noise $z_t$ and is given by $\bar{O}=F(t)J_-+G(t)J_zJ_-+P_z(t) J_-^2$, where $F(t)$, $G(t)$ and $P_z(t)$ are time-dependent functions that can in principle be calculated once the correlation function $\alpha(t,s)$ is known. {It is clear that we can assume an upper-triangular invariant of the form}
\begin{equation}
	I(t)=
	\begin{pmatrix}
	0&a(t)&b(t)\\
	0&1&c(t)\\
	0&0&2
	\end{pmatrix}.
\end{equation}
{Using the commutation relationship of the ladder operator for the three level system, we found from the definition Eq.~\eqref{i_heff_def}} the dynamic invariant $I(t)$ for this model
\begin{align}
	&a(t)=\mathcal{R}\left[2(F(t)+G(t))-i \omega,\sqrt{2}z_t\right], \nonumber\\
	&b(t)=\mathcal{R}\left[F(t)-i \omega,\sqrt{2} \left(2a(t)P_z(t)-\left(a(t)-c(t)\right)z_t\right) \right], \nonumber\\
	&c(t)=\mathcal{R} \left[-2G(t)-i \omega,\sqrt{2}(z_t-2P_z(t))\right].
\end{align}
where $\mathcal{R}[g(t),h(t)]=\int_0^t \exp \left(\int_u^t g(s)ds\right)h(u)du$.
We then have
\begin{align}
	|\psi_{z^*}(0)\rangle&=\psi_1(0)e^{-i \omega t}|0\rangle \nonumber\\
	&+\psi_2(0)\exp\left[-2\int_0^t F(s)+G(s)ds\right]|\varphi_2(t) \rangle \nonumber\\
	&+\psi_3(0)\exp\left[i \omega t-2\int_0^t F(s)ds\right]|\varphi_3(t)\rangle,
\end{align}
where 
\begin{align}
	|\varphi_2(t) \rangle&=a(t)|0 \rangle+|1 \rangle, \nonumber\\
	|\varphi_3(t) \rangle&=\frac{b(t)+a(t)c(t)}{2}| 0 \rangle+c(t) |1 \rangle+|2 \rangle.
\end{align}

\begin{figure}[t]
	\begin{centering}
		\includegraphics[scale=.4]{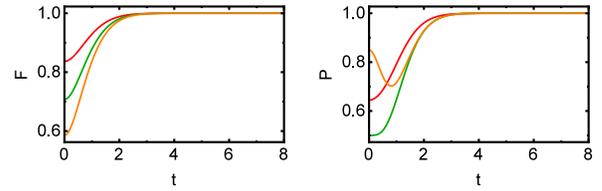}
	\end{centering}
	\caption{(Color online) The target fidelity and purity of three random initial states marked by {dashed} red, {dotted orange} and {solid} green lines as a function of time. Under the Hamiltonian we reversely engineered, the initial states are monotonically driven to the target pure steady-state by non-Markovian dynamics.}
	\label{fig_re}
\end{figure}

It analytically reveals the quantum trajectory of the dissipative three-level system.

\section{Reverse engineering}
Now we show how to use dynamical invariants to design a Hamiltonian that can be used to drive an initial state to a target state by means of reverse engineering~\cite{Jing2013b}. Specifically, to design the needed Hamiltonian, we first construct an invariant $I(t)$ such that one of its time-dependent eigenvector $|\varphi_1 \rangle$ follows the desired time-evolution path, according to Eq.~\eqref{psi_solu}. The rest eigenvectors $\mathcal{R}=\{| \varphi_i\rangle\}$, $i=2,\ldots,N$, are left as undetermined parameters which we will use later to make the invariant compatible with the QSD equation. Then, we take the time-derivative of this invariant to obtain its equation of motion, and compare it with Eq.~\eqref{i_heff_def}, where $H_{\rm eff}$ should be formally compatible with Eq.~\eqref{heff_qsd_def}, imposed by the QSD equation, i.e., $H_s$ should be hermitian, both $H_s$ and $L$ should be \emph{noise-independent}, and the $\bar{O}$ operator is determined by $H_s$ and $L$. This is achievable by choosing an appropriate set of basis $\mathcal{R}$. We then have the desired Hamiltonian $H_s$ of the system and the corresponding $L$ operator.

As an illustrative example, we consider a two-level open system with the target state $|\psi_T \rangle=(|0 \rangle+|1 \rangle)/\sqrt{2}$. By letting this state be one of the eigenvector of the invariant, we first make a noise-dependent invariant of the form
\begin{equation}
	I(t)=\begin{pmatrix}p(t,z^*_t)&-p(t,z^*_t)-1\\p(t,z^*_t)-1&-p(t,z^*_t)\end{pmatrix},\label{i_re}
\end{equation}
where $p(t,z^*_t)$ is a function determined by the Hamiltonian $H_s$ of the system and Lindblad operator $L$ that we are designing. If the coefficient $c_\mu(t)$ in the general solution to Eq~\eqref{psi_solu} decays to zero for the other eigenvector of the invariant, we certainly find the steady target state. Unlike a closed system, the specific form of the effective Hamiltonian of the QSD poses constraints on $H_s$ and $L$ in that $H_s$ needs to be hermitian and noise-independent. With both $H_s$ and $L$ given, the $\bar{O}$ operator is then determined. Taking this into consideration and inserting Eq.~\eqref{i_re} into Eq.~\eqref{i_heff_def}, we find that $H_s=-\omega \sigma_x$ and $L=\lambda(\sigma_z-i \sigma_y)$. In Fig.~\ref{fig_re} we numerically plot the fidelity between the engineered state and the target state, as well as the purity of the engineered state for three randomly chosen mixed initial states. It can be seen that the fidelity monotonically increases to one and become unity after some time, indicating a steady target state is reached. In sharp {contrast} to the closed quantum system that evolves unitarily, the spectrum of the state is free to change, and we can drive a \emph{mixed} state to a target \emph{pure} state by using the non-Markovian dynamics of an open quantum system.

\section{Conclusion}
In conclusion, we studied the dynamical invariants in non-Markovian open systems whose temporal evolution is governed by the non-Markovian QSD equation. {For systems that have an exact operator representation of the function derivative in the form of the $O$ operator, the dynamical invariant can be obtained analytically.} Dynamical invariants of the QSD equation are discovered, and it is found that the non-hermitian dynamical invariants do not share the same equation of motion as the reduced density matrix but its eigenvectors can be used to generate an analytical expression of the solution of QSD. This enables us to obtain temporal evolution of the open system without deriving and then solving the non-Markovian master equations. Using reverse engineering along with the QSD invariants, we are able to design a Hamiltonian and Lindblad operator that can be used to drive an initial state to a target state via non-Markovian evolution.

\appendix*
{\section{Derivation for the solution to QSD}}
{Using Eq.~\eqref{i_heff_def}, we take the time derivative of
\begin{align}
	I|\varphi_\lambda\rangle&=\lambda|\varphi_\lambda\rangle,\label{ieig}
\end{align}
and project it onto $\langle\widetilde{\varphi}_\mu|$,
\begin{align}
	&\quad -i\lambda\langle\widetilde{\varphi}_\mu|H_{\rm eff}|\varphi_\lambda\rangle+i\mu\langle\widetilde{\varphi}_\mu|H_{\rm eff}|\varphi_\lambda\rangle+\mu\langle\widetilde{\varphi}_\mu|\dot{\varphi}_\lambda\rangle \nonumber \\
	&=\dot{\lambda}\langle\widetilde{\varphi}_\mu|\varphi_\lambda\rangle+\lambda\langle\widetilde{\varphi}_\mu|\dot{\varphi}_\lambda\rangle.
\end{align}
Thus,
\begin{align}
	\dot{\lambda}\delta_{\mu,\lambda}&=\left(\mu-\lambda\right)\left[\langle\widetilde{\varphi}_\mu|\dot{\varphi}_\lambda\rangle+i\langle\widetilde{\varphi}_\mu|H_{\rm eff}|\varphi_\lambda\rangle\right],\\
	\mu=\lambda&\Rightarrow \dot{\lambda}\equiv 0,\\
	\mu\neq \lambda &\Rightarrow \langle\widetilde{\varphi}_\mu|\dot{\varphi}_\lambda\rangle=-i\langle\widetilde{\varphi}_\mu|H_{\rm eff}|\varphi_\lambda\rangle.\label{ieig_hsol}
\end{align}
We then expand the wave function in this basis:
\begin{align}
	|\psi(t)\rangle=\sum_\lambda c_\lambda(t)|\varphi_\lambda(t)\rangle.
\end{align}
Inserting it into the QSD equation and projecting onto $\langle\widetilde{\varphi}_\mu|$, we have
\begin{align}
	\sum_\lambda \dot{c_\lambda}\delta_{\mu,\lambda}+\sum_\lambda c_\lambda\langle\widetilde{\varphi}_\mu|\dot{\varphi}_\lambda\rangle&=-i\sum_\lambda c_\lambda \langle\widetilde{\varphi}_\mu|H_{\rm eff}|\varphi_\lambda \rangle.
\end{align}
Therefore,
\begin{align}
	\dot{c_\mu}&=\sum_\lambda \left[-i c_\lambda \langle\widetilde{\varphi}_\mu|H_{\rm eff}|\varphi_\lambda \rangle-c_\lambda\langle\widetilde{\varphi}_\mu|\dot{\varphi}_\lambda\rangle\right],
\end{align}
using Eq.~\eqref{ieig_hsol},
\begin{align}
	\dot{c_\mu}&=\sum_{\lambda\neq\mu} \left[-i c_\lambda \langle\widetilde{\varphi}_\mu|H_{\rm eff} |\varphi_\lambda \rangle+ic_\lambda\langle\widetilde{\varphi}_\mu|H_{\rm eff}|\varphi_\lambda\rangle\right] \nonumber\\
	&\phantom{=} -i c_\mu \langle\widetilde{\varphi}_\mu|H_{\rm eff} |\varphi_\mu \rangle-c_\mu\langle\widetilde{\varphi}_\mu|\dot{\varphi}_\mu\rangle \nonumber\\
	&=-c_\mu \left[i \langle\widetilde{\varphi}_\mu|H_{\rm eff} |\varphi_\mu \rangle+\langle\widetilde{\varphi}_\mu|\dot{\varphi}_\mu\rangle\right]
\end{align}
Now the differential equations for coefficients $c_\mu(t)$ are decoupled and can be readily solved,
\begin{align}
	c_\mu(t)=c_\mu(0)\exp \left(-\int_0^t d\tau \left[i \langle\widetilde{\varphi}_\mu|H_{\rm eff} |\varphi_\mu \rangle+\langle\widetilde{\varphi}_\mu|\frac{\partial}{\partial \tau}|\varphi_\mu\rangle \right]\right)
\end{align}
}

\begin{acknowledgments}
This work is supported by the Basque Government (Grant No.~IT472-10), the Spanish MICINN (Project No.~FIS2012-36673-C03-03), and the Basque Country University UFI (Project No.~11/55-01-2013). C.H.L. is supported by the Hong Kong GRF (Project No.~501213). J. Q. You is supported by the National Natural Science Foundation of China No.~91421102 and the National Basic Research Program of China No.~2014CB921401. T.Y. is supported by the NSF PHY-0925174 and DOD/AF/AFOSR No.~FA9550-12-1-0001. We would like to thank J. G. Muga for helpful discussions.
\end{acknowledgments}



\end{document}